\documentclass[prb,twocolumn,superscriptaddress,showpacs]{revtex4}

\usepackage{graphicx}

\newcommand{\EF}{$E_{\rm F}$}
\newcommand{\kF}{$k_{\rm F}$}

\newcommand{\LPB}{Li$_{0.9}$Mo$_6$O$_{17} $}
\newcommand{\hv}{$h\nu$}

\begin{document}

\title{The case for the bulk nature of the spectroscopic Luttinger liquid signatures

observed in angle resolved photoemission of \LPB}


\author{Feng Wang}
\author{S.-K. Mo}
\author{J. W. Allen}
\affiliation{Randall Laboratory of Physics, University of
Michigan, Ann Arbor, MI 48109, USA}

\author{H.-D. Kim}
\affiliation{Pohang Accelerator Laboratory, Pohang 790-784, Korea}

\author{J. He}
\affiliation{Department of Physics and Astronomy, University of
Tennessee, Knoxville, Tennessee 37996, USA}
\author{R. Jin}
\affiliation{Materials Science and Technology Division, Oak Ridge
National Laboratory, Oak Ridge, Tennessee 37831, USA}
\author{D. Mandrus}
\affiliation{Department of Physics and Astronomy, University of
Tennessee, Knoxville, Tennessee 37996, USA} \affiliation{Materials
Science and Technology Division, Oak Ridge National Laboratory,
Oak Ridge, Tennessee 37831, USA}

\author{A. Sekiyama}
\author{M. Tsunekawa}
\author{S. Suga}
\affiliation{Department of Material Physics, Graduate School of
Engineering Science, Osaka University, 1-3 Machikaneyama,
Toyonaka, Osaka 560-8531, Japan}

\date{\today}

\begin{abstract}
Angle resolved photoemission spectroscopy (ARPES)
has been performed on quasi-one dimensional \LPB\ using photon
energy $h\nu = 500$ eV.  Measured band dispersions are in
agreement with those from both low photon energy measurements and
band structure calculations.  The momentum integrated ARPES
spectrum is well fit by the finite temperature Luttinger liquid
(LL) spectral function, with an anomalous exponent 0.6 that is the
same within experimental uncertainty as the value found with $h\nu
= 30$ eV.  These identical findings at both low and high \hv\ are entirely consistent with reasoning based on the crystal structure, that the quasi-one dimensional chains lie two layers below the cleavage plane so that the observed spectroscopic LL behavior of \LPB\ is a bulk property.

\end{abstract}


\maketitle

The low energy physics of a 1-dimensional (1D) interacting
electron system is described by the Luttinger liquid (LL)
picture~\cite{Voit94}, in which there are no single particle
excitations, but only collective modes of the charge and spin
density, holons and spinons, respectively.  In approaching the
Fermi energy (\EF), the energy dependence of the momentum-summed
single particle density of states (DOS) displays a power law decay
with an anomalous exponent, i.e. $|E|^{\alpha}$.  This Brief
Report adds an important element to the spectroscopic
evidence presented in previous work
~\cite{JDD99,Allen02,Gweon01,Gweon02,Gweon03,Gweon04,Wang06} that
\LPB\ provides a paradigm solid material for the study of LL
physics.

\begin{figure}[!t]
\includegraphics[width=3.3 in]{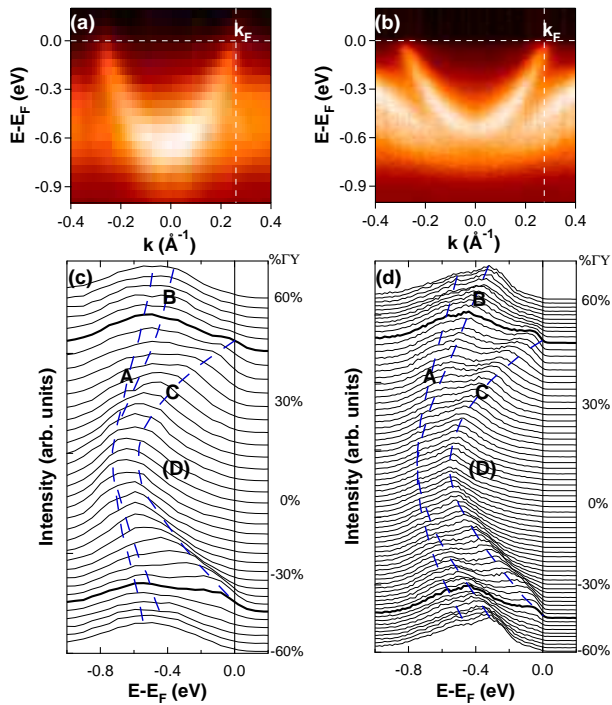}
\caption{(Color online) (a),(b): Energy $vs.$ momentum spectra
along $\Gamma (k = 0)$ -Y direction at \hv\ = 500 eV and 30 eV,
with resolutions $\Delta E \approx$ 180~meV and 18~meV,
respectively. The color represents the spectral intensity, with
white for the highest and black for the lowest. The image in (b)
is symmetrized from data taken between $k = 0 - 0.4$ \AA$^{-1}$.
(c),(d): Stack view of panel (a) and (b), with the thick lines
highlighted for $k = k_{F}$, within momentum resolution. The
dashed lines in (c) and (d) highlight the band dispersions. Band D
is suppressed by matrix element in this geometry and is indicated
by parenthesis.}
\end{figure}

\LPB ~is a quasi-1 dimensional metal having highly anisotropic
electronic properties \cite{Greenblatt84,Choi04}. We have studied
this material extensively by angle resolved photoemission
spectroscopy
(ARPES)~\cite{JDD99,Allen02,Gweon01,Gweon02,Gweon03,Gweon04,Wang06}.
The momentum dependent spectra reveal a band structure that is in
good general agreement with that of tight-binding band
calculations\cite{Whangbo88}, and detailed comparison to LL
spectral theory \cite{Orgad01} shows that the ARPES lineshape has
expected holon and spinon features moving with different
velocities.  As is then expected, and of direct interest for this
paper, the momentum integrated spectrum shows a clear power law
suppression of the near \EF\ DOS.  The power law DOS has also been
observed in scanning tunnelling spectroscopy\cite{Hager05}. Most
recently we have found~\cite{Wang06} that $\alpha$ shows a strong
temperature (T) dependent renormalization such that the values
found in tunneling and ARPES are quantitatively consistent, and
that this renormalization is the result of marginal interactions
among charge neutral modes present explicitly because of the two
band nature of \LPB.

\begin{figure*}
\includegraphics[width=6 in]{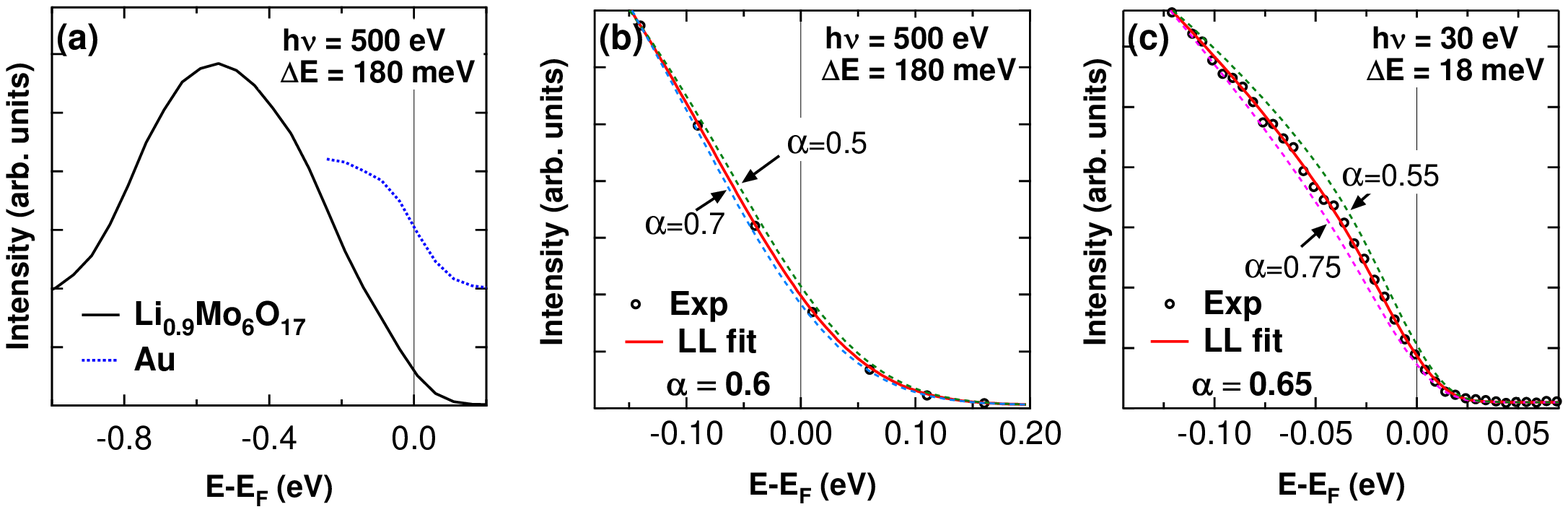}
\caption{(Color online) (a): Momentum integrated spectra of panel
1(a). (b), (c): Fit of momentum integrated spectra with finite T
LL theory. The solid lines in (b) and (c) are the best fit and the
dashed lines are calculations with $\alpha$\ slightly off the
optimal value, as indicated by the arrows.}
\end{figure*}

The new spectroscopic element presented here is to show that ARPES
data obtained at high photon energy (\hv $\geq500$ eV) yield the
same value of $\alpha$ as obtained for the ARPES performed to date,
done with much lower photon energy (\hv\ $\leq 30$ eV). This finding
bears on the case that the LL properties found in low \hv\ ARPES and
tunneling spectroscopy are characteristic of the bulk solid.  The
case is already strong, because, from the crystal structure, the
likely cleavage plane is separated from the quasi-1D conducting
chains by 4-8 \AA\cite{Onoda87}, such that even the chains nearest
the surface are shielded from the external environment by two layers
of MoO$_6$ octahedra and MoO$_4$ tetrahedral, and locally have the
bulk environment.  Thus, even though the value of the low energy
electron mean free path (EMFP) is \textit{a priori} very uncertain
because it depends on the details of low energy excitations of a
material, and could be as small as $<5$\AA, it is highly plausible
that the chain electronic structure measured in tunneling and low
\hv\ ARPES is a bulk property.  Indeed, no aspect of these data has
suggested otherwise.  Nonetheless the importance of the issue
motivated further testing, here by using increased \hv\ to increase
the likelihood of a large EMFP through higher kinetic energy
photoelectrons.

The EMFP of electrons with large kinetic energy ($>150$ eV) is much
less affected by the details of low energy excitations and
characteristically increases monotonically with the kinetic
energy\cite{Powell99}.  Published data\cite{Powell99} typically show
EMFP's to be larger at 500 eV than at 30 eV by factors of 2 and even
more.  Although energy and momentum resolutions available in high
\hv\ ARPES are still not as good as those available at low \hv,
recent improvements~\cite{Saitoh00} in synchrotron radiation and
electron detection technologies have made it
possible~\cite{Sekiyama04a,Suga04} to obtain spectra that can be
meaningfully compared to low \hv\ spectra.  As a result, examples
are now known where spectra using high \hv\ similar to this report
have revealed new features~\cite{Suga04, Mo03} not found at low \hv.
Some of these examples are low dimensional materials~\cite{Suga04}.
Although it is not possible to be quantitative as to the increase in
EMFP for \LPB\ from 30 eV to 500 eV, it is certain that differing
ARPES results for the two \hv\ values would signal a serious problem
with the case for bulk LL spectroscopic signatures based on the
reasoning from the crystal structure set forth above.  However, as
described below, no such difference is found.

High \hv\ ARPES measurements were made at the twin-helical undulator
beam line BL25SU of SPring-8\cite{Saitoh00}, using circularly
polarized photons with energy 500~eV.  The beamline is equipped with
a SCIENTA SES200 electron kinetic energy analyzer. Single crystal
\LPB\ samples grown using the temperature gradient flux method were
oriented by Laue diffraction so that the 1D conducting direction
($b$ axis) was aligned to the analyzer slit direction. The samples
were cleaved \textit{in situ} in a vacuum better than $2 \times
10^{-10}$~Torr to expose clean surfaces of the $ab$-plane. The Fermi
level and overall energy resolution ($\Delta E \approx$ 180~meV)
were determined from a Fermi edge spectrum taken on freshly
evaporated Au. The momentum resolution was about 0.05 \AA$^{-1}$.
The temperature was controlled by an embedded resistive heater and a
closed-cycle He cryostat. Photon-induced sample damage, which
appears as an extra peak around 2~eV binding energy, was observed
after 5 - 6 hours of measurement.  We carefully chose the experiment
parameters so that all the data presented in this paper were taken
within 3 hours of the time that a freshly cleaved surface was newly
exposed to the light. For example, our use of a 50 meV energy step
size was deemed to be a good balance between reducing the time to
take data and yet having adequate point density for the data
analysis presented below. Measurements with linearly polarized
photons at \hv\ = 30 eV were performed at the Synchrotron Radiation
Center of the University of Wisconsin, with similar angle resolution
and thus somewhat better momentum resolution, but with $\Delta E
\approx$ 18~meV, and with other experimental details the same as in
Ref.~[\onlinecite{Wang06}].

Fig. 1(a) and (b) show a comparison of the energy $vs.$ momentum
({\it k}) spectra measured at \hv~= 500 eV and \hv~= 30 eV,
respectively. Both spectra are measured along the $\Gamma$-Y
direction and at T = 50 K. \LPB\ has 4 Mo 4{\it d} bands near \EF,
labeled as A-D  in the stack plots of Fig. 1(c) and 1(d).  The 4th
band (D), which has the lowest binding energy and disperses to merge
with band C before crossing \EF, is known to have a greatly
suppressed intensity along this k-path and hence is not seen. This
band is observable~\cite{Wang06} along another path in k-space that
was not readily accessible for the experimental setup used here. The
spectra obtained using the two different photon energies are
generally very similar to each other and to the results of band
structure calculations~\cite{Whangbo88}. Differences between the two
sets of data arise from the differing resolutions and also from
differing photoelectron matrix elements, including perhaps the
effect of the differing photon polarizations,  that highlight the
A-B bands differently. For example, in Fig. 1(a) the high intensity
centered at binding energy $\sim0.7$ eV and $k = 0$ is from the A-C
bands not well resolved. The band minimum of C is actually at $\sim
0.55$ eV, as becomes clear in Fig. 1(b), where the A and B bands are
weaker near $\Gamma$ and the energy resolution is much better.  The
Fermi vector extracted from the 500 eV spectra is 2\kF\ = 0.52
\AA$^{-1}$~compared to 2\kF\ = 0.55 \AA$^{-1}$~obtained at 30 eV
photon energy, adequate agreement within the momentum resolution.

The spectra in Fig. 1 were k-integrated to obtain the spectra
shown in Fig. 2.   Fig. 2(a) shows the full spectrum corresponding
to the \hv~= 500 eV data of Fig. 1(a).  Compared to the gold Fermi
edge measured for the same experimental conditions and also shown
in the figure, it is clear that there is a suppression of near
\EF\ spectral weight.  Fig. 2(b) shows the near \EF\ spectrum for
\hv~= 500 eV on an expanded energy scale, with the data plotted as
circles, and for comparison, Fig. 2(c) shows the near \EF\
spectrum corresponding to the \hv~= 30 eV data of Fig. 1(b). The
former shows a noticeable broadening relative to the latter due to
the relatively poorer energy resolution for \hv~= 500 eV.

To quantitatively characterize the k-integrated lineshapes taken
at the two photon energies, we compare them with the theoretical
LL DOS for finite T from Ref. [\onlinecite{Orgad01}]. It is very
important to account for the differing experimental energy
resolutions, by convolving the calculated lineshape for T = 50 K
with a Gaussian of width set to the appropriate resolution $\Delta
E$. For \hv~= 500 eV we find that the theory lineshape fits the
spectrum very well up to binding energy 0.14 eV with $\alpha =
0.6$ as shown in Fig. 2(b). The same analysis of the \hv~= 30 eV
spectrum gives $\alpha = 0.65$ with a fitting range up to 0.12 eV
binding energy, as shown in Fig. 2(c). As expected from this
analysis, direct broadening of the \hv~= 30 eV spectrum (not
shown) yields a curve differing only slightly from the \hv~= 500
eV spectrum. The slightly different values of $\alpha$ are within
expectations either from the degree of sample dependence observed
previously~\cite{Wang06}, or from the fitting uncertainty,
estimated to be $\sim \pm0.05$. Both alpha values extracted are
also in very good agreement with the scanning tunneling
result~\cite{Hager05}, which gives $\alpha\ = 0.62\pm 0.17$ at 5
K, considering that the T dependent study\cite{Wang06} shows very
little variation of $\alpha$ for temperatures below 50K. Thus we
conclude that the momentum integrated spectra at both photon
energies reflect the same underlying LL spectrum.


In summary, we have presented the results of bulk sensitive ARPES
spectroscopy on \LPB\ performed with \hv~= 500 eV. The measured band
structure and an analysis of the momentum integrated spectrum are in
good agreement with the results of ARPES experiments using low
photon energies.  This finding is fully consistent with our
reasoning based on the crystal structure,  that the chains lie well
below the cleavage plane and hence are well shielded from the
effects of the surface, such that the LL ARPES lineshapes observed
for \LPB\ are a bulk property.

\begin{acknowledgments}
This work was supported at UM by the U.S. NSF (DMR-03-02825), at
UT by the U.S. NSF (DMR 00-72998), by Office of Basic Energy
Sciences, U.S. DOE (DE-AC05-00R22725) at ORNL, managed by
UT-Battelle, LLC., by the Center for Strongly Correlated Materials
Research, Korea, by Grant-in-Aid for COE Research(10CE2004) and
Grant-in-Aid for Creative Scientific Research(15GS0213) of MEXT,
Japan, by 21st Century COE Program(G18) by Japan Society for the
Promotion of Science.
\end{acknowledgments}

\end{document}